\documentclass[twocolumn]{aastex631x}

\newcommand{\kepler}{\textit{Kepler}}
\newcommand{\ktwo}{\textit{K2}}
\usepackage{xcolor}
\usepackage{hyperref}

\graphicspath{{./}{figures/}}

\begin{document}

\title{A Phenomenon Resembling Early Superhumps in a New SU UMa-Type Dwarf Nova with a 2-Hour Orbital Period}

\author{Rebecca S. Boyle}
\affiliation{Department of Physics, University of Notre Dame, Notre Dame, IN 46556, USA}
\author[0000-0001-7746-5795]{Colin Littlefield}
\affiliation{Bay Area Environmental Research Institute, Moffett Field, CA 94035 USA}
\affiliation{Department of Astronomy, University of Washington, Seattle, WA 98195, USA}
\author{Peter Garnavich}
\affiliation{Department of Physics, University of Notre Dame, Notre Dame, IN 46556, USA}
\author{Ryan Ridden-Harper}
\affiliation{School of Physical and Chemical Sciences | Te Kura Mat\={u}, University of Canterbury,
Private Bag 4800, Christchurch 8140,\\ Aoteroa, New Zealand}
\author[0000-0003-4373-7777]{Paula Szkody}
\affiliation{Department of Astronomy, University of Washington, Seattle, WA 98195, USA}
\author{Patricia Boyd}
\affiliation{NASA Goddard Space Flight Center, Greenbelt, MD 20771, USA}
\author{Krista Lynne Smith}
\affiliation{Department of Physics, Southern Methodist University, Dallas, TX 75205, USA}

\begin{abstract}

We investigate K2BS5, an optical transient that we identified in Campaign~13 of the \kepler/\ktwo\ archives by the \ktwo\ Background Survey, and classify it as a new SU~UMa-type dwarf nova. Using the light curve generated from \kepler’s long-cadence observation mode, we analyze the dwarf nova during quiescence and superoutburst. Following 20 days of quiescence at the start of the observation, the system entered a superoutburst lasting 12 days, after which it experienced at least one rebrightening. K2BS5 clearly meets the criteria for an SU UMa star, but at the peak of the superoutburst, it also shows double-wave oscillations consistent with the spectroscopic orbital period, a phenomenon that closely resembles early superhumps in WZ~Sge stars. While we do not classify K2BS5 as a WZ~Sge system, we discuss how this phenomenon could complicate efforts to use the suspected detection of early superhumps to distinguish SU~UMa-type dwarf novae from the recently recognized class of long-orbital-period WZ~Sge systems.

\end{abstract}


\section{Introduction} \label{sec:intro}

\begin{figure*}
    \centering
    \includegraphics[width=\textwidth]{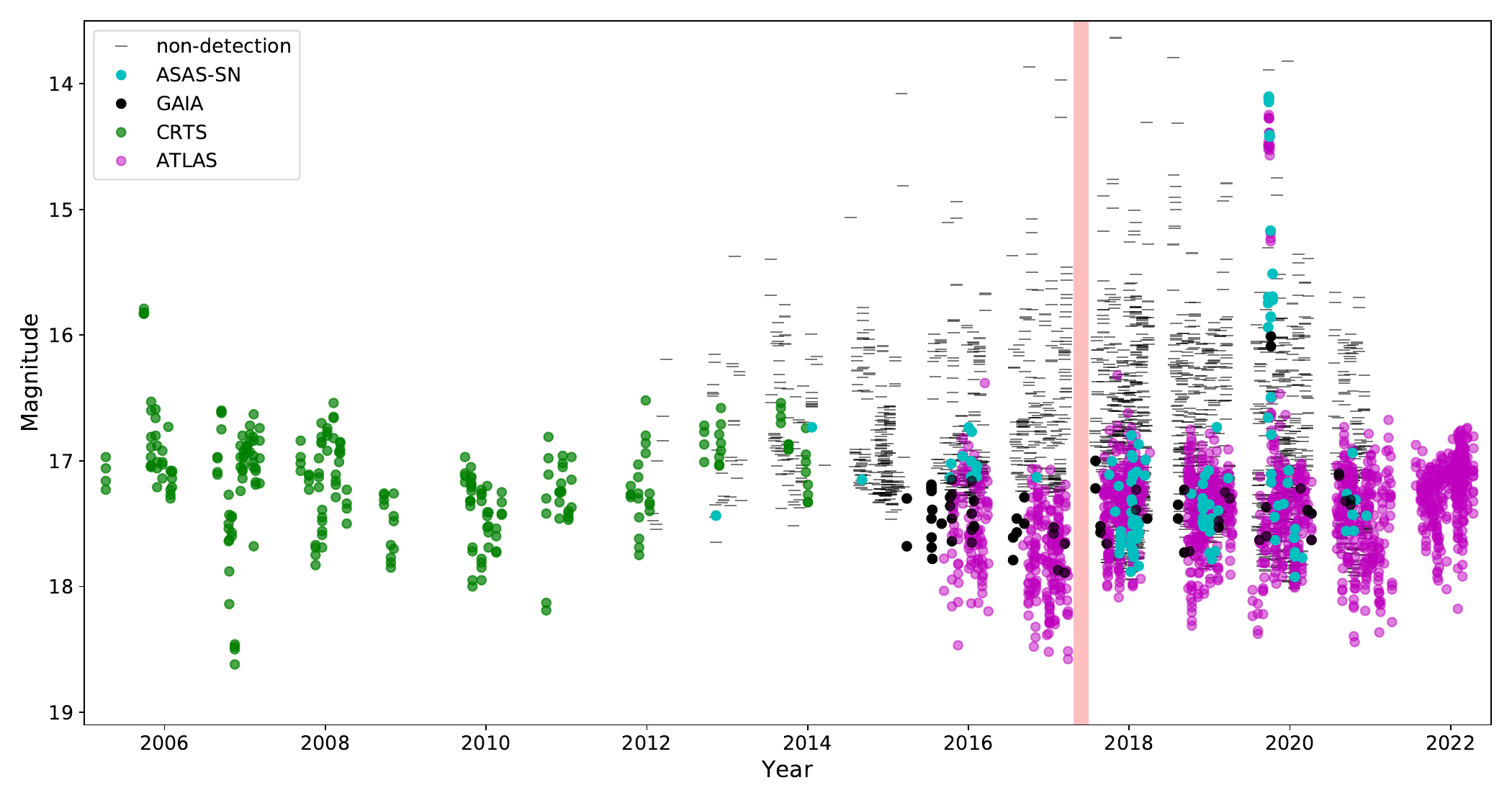}
    \caption{The ASAS-SN, Gaia, CRTS, and ATLAS extended lightcurves from February 2005 to April 2022.
    The more sparse sampling shows evidence of few notable events apart from one superoutburst occurring in September 2019. There are no normal outbursts beyond superoutbursts, a hallmark of WZ Sge systems. The superoutburst observed by  \kepler\, whose observation time is highlighted in red, is not visible because the system was near solar conjunction at the time. 
    }

    \label{fig:longterm_light_curve}
\end{figure*}

    Cataclysmic variables (CVs) are a classification of binary star systems consisting of a white dwarf (WD) primary paired most commonly with a red dwarf (RD) secondary. Mass transfer between the two stars occurs when the secondary overflows its Roche lobe, resulting in the formation of an accretion disk around the primary if the WD is not strongly magnetized \citep[for reviews, see][]{warner95, hellier_2001}. Often the accretion disk is thermally unstable and subject to recurring outbursts on timescales ranging from days to many years, depending on the mass-transfer rate \citep[][and references therein]{osaki_review}. These systems are known as dwarf novae \citep[DN; for a review, see][]{osaki_review}.
    
    SU UMa systems are a subcategory of DN with typical orbital periods $\lesssim$ 2~h that are distinguished by the occasional ``superoutbursts'' they experience, which are outbursts of longer duration and greater amplitude in comparison to ordinary outbursts. During superoutbursts, SU~UMa systems show the presence of superhumps, which are periodic oscillations slightly below the orbital frequency. Superhumps result from tidal instability in the accretion disk, excited when the outer disk expands to the 3:1 resonance with the orbital period of the binary \citep{whitehurst88, osaki}. As seen in \citet{Kato2009} and \citet{Kato_2022_V844}, the superoutburst displays three distinct phases of period evolution, designated as stages A, B, and C. Stage A appears first in the evolutionary progression, characterized by the longest superhump period and no discernible period derivative. Stage B is the middle segment with a positive period derivative, followed by stage C exhibiting a shorter and more stable period.

    WZ Sge-type systems \citep[reviewed by][]{Kato2015} are a subcategory of the SU UMa-type that generally show only superoutbursts whose recurrence times are significantly longer than those of SU~UMa stars. Like the SU UMa systems, they exhibit long-duration superoutbursts,  
    but they can often be distinguished by the presence of subsequent rebrightening events and low-amplitude oscillations in the earliest stages of the superoutburst. The rebrightenings are also referred to as echo outbursts.
    %believed to be the result of continued tidal distortion in the accretion disk after the conclusion of the superoutburst \citep{Hellier_Paper}. 
    The double-wave feature is referred to in the literature as early superhumps (ESH), which have approximately the same period as the binary orbital period. ESH are thought to be the result of expansion of the outer accretion disk to a 2:1 resonant frequency with the orbital period, which is possible only for very small mass ratios. The subsequent transition from ESH into Stage A takes place as the outer disk at the 3:1 resonance radius becomes eccentric and undergoes apsidal precession \citep{Kato2013}.
    
    Observing these short-lived phenomena during superoutbursts has proven to be one of the many accomplishments of the \kepler\ spacecraft during its original and \ktwo\ missions, both of which provided continuous light curves of predetermined targets for an extended period of time (often months or longer). However, while previous analyses of \kepler\ data have focused on known CVs, the \kepler\ archives contain numerous background pixels that were not studied directly at the time of observation, providing an opportunity to search for previously unknown, interesting objects. Within those background pixels, the \ktwo\ Background Survey \citep[K2BS;][]{RH2020} has uncovered transients by systematically identifying potential transients that are then reviewed manually. One of the early successes of the K2BS project was the discovery of the only superoutburst of a WZ~Sge system observed by \kepler /\ktwo\ mission \citep{RH2019}.
  
    Here we present a photometric and spectral analysis of K2BS5, a new SU UMa-type DN.

\section{Data} \label{sec:data}

\begin{figure*}
    \centering
    \includegraphics[width=\textwidth]{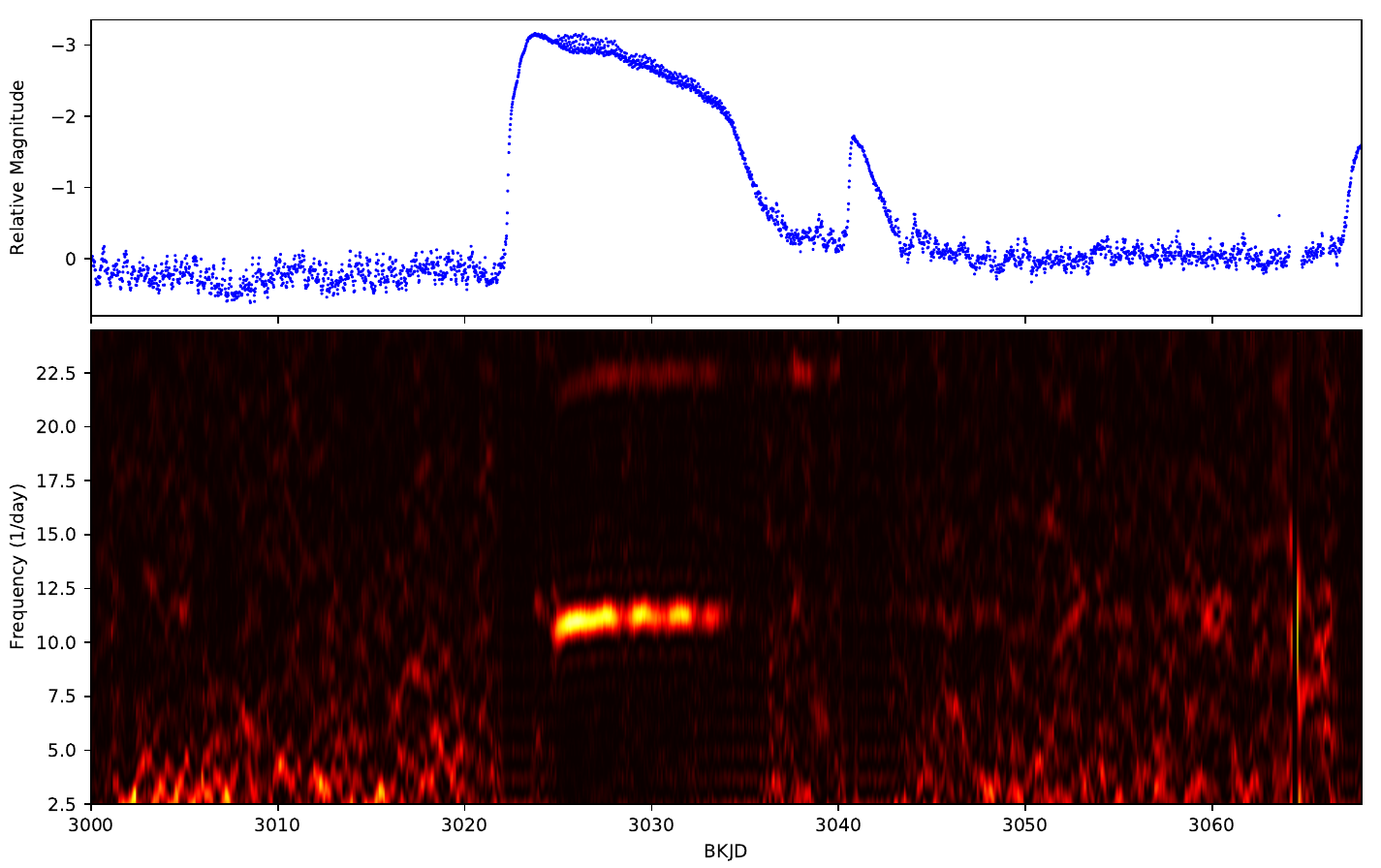}
    \caption{The full \kepler/\ktwo\ light curve shown in magnitudes relative to the quiescent magnitude (top panel) and time-resolved power spectrum of K2BS5 (bottom panel). The light curve shows a long-period of quiescence before the steep rise at the beginning of the superoutburst. A decrease in magnitude near the end of the superoutburst precedes an \textbf{rebrightening event} before the system resumes its original state of quiescence. The time-resolved power spectrum shows that periodic variability is present only during the superoutburst.}
    \label{fig:Superoutburst}
\end{figure*}

     Unlike the serendipitous background sources that the K2BS project normally seeks to unearth, K2BS5 was targeted in \ktwo\ Campaign~13 as part of a program (G013086, P.I. Patricia Boyd) to observe candidate active galactic nuclei identified from a search of archival X-ray sources. In addition to the Chandra detection that led to its inclusion in that \ktwo\ program, it is listed in the 2XSPS Catalog of Swift X-ray Telescope Point Sources and Data Release 8 of the XMM-Newton Serendipitous Source Catalog; its identifiers in these three catalogs are CXOJ043436.4+180243, 2SXPS~J043436.5+180243, and 3XMM~J043436.6+180245, respectively. We refer to the object here as K2BS5 because it was the fifth transient discovered by the K2BS project, but it has also been detected by the All Sky Automated Survey for SuperNovae \citep[ASAS-SN;][]{shappee, kochanek} under the designation ASASSN-19xs. It has two additional identifiers (Gaia19emm and AT~2019sgc) as a result of the detection by Gaia of an outburst in 2019 \citep{2019TNSTR2030....1H}.

    K2BS5 is an under-studied system. The only significant attention that it has received was a spectroscopic measurement of its orbital period ($123.55\pm0.09$~min)  \citet{thorstensen20}, who used the identifier Gaia19emm. Its Gaia EDR3 \citep{gaia_edr3} position is $\alpha_{2000} = 04\rm{h}34\rm{m}36.6053\rm{s}, \delta_{2000} = +18^{\circ}02\rm{'}45.025\rm{"}$. \citet{Bailer-Jones} determined the system's distance to be 472~$\pm$~31~pc based on the Gaia EDR3 parallax.

\subsection{\ktwo\ light curve}

During \ktwo, \kepler\ suffered from a 6~hr periodic drift, causing targets to shift across the detector and degrading the photometric precision. For well-isolated targets, one way of mitigating this problem is to select a sufficiently large photometric aperture that all of the target's flux is captured, regardless of the drift motion. Using the interactive inspection tool in {\tt lightkurve} \citep{Lightkurve}, we selected a custom extraction aperture from the target pixel file data to encompass the system's full range of motion over the course of the observation. Additionally, we used {\tt lightkurve} to visually inspect the images of the source in order to confirm that its brightness variations were attributable to variations in the target (and not from spacecraft systematics or the passage of an asteroid through the photometric aperture). The data were obtained during the \kepler/\ktwo\ Campaign 13, running from 2017 March 8 until 2017 May 27, a duration of $\approx$ 2.5 months, at a 30 minute cadence.

\subsection{LBT Spectrum}

We obtained spectra of K2BS5 with the Multi-Object Dual Spectrograph \citep[MODS;][]{pogge12} on the Large Binocular Telescope (LBT). Nine individual spectra were obtained on 2020 February 29 (UT) under cloudy conditions. The final three spectra had the strongest signal, and these were averaged to create the final spectrum with a total exposure time of 900s. The dual grating mode for MODS was combined with a 0.8~arcsec slit to provide a spectral resolution of $R=1860$ at H$_\beta$. Seeing during the exposures varied between 1.1 and 1.5 arcsec.

The spectra were extracted and wavelength calibrated using argon and neon emission line arcs. The spectra were flux calibrated using the spectrophotometric standard star Feige~34 obtained on a clear night earlier in the run. 

\subsection{Krizmanich Photometry}
We conducted additional ground-based observations of K2BS5 on four nights during the first week of March 2021 using the University of Notre Dame's 0.8m Sarah L. Krizmanich Telescope (SLKT). The time was corrected to Barycentric Julian Date \citep{Eastman} with {\tt astropy} \citep{Astropy}. During each of the 2~hour long observations, we obtained unfiltered images using 30 second exposures. The typical signal to noise ratio per exposure was approximately 14.0.

\begin{figure}
    \centering
    \includegraphics[width=\columnwidth]{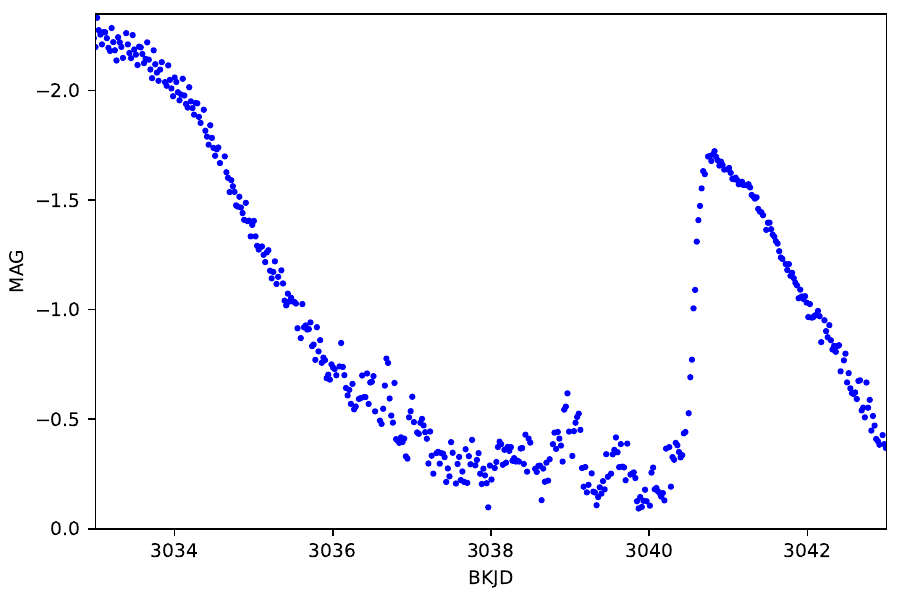}
     \caption{Low-amplitude oscillations appear during the post-superoutburst phase just before the \textbf{rebrightening}. These $\sim$0.4~mag oscillations are reminiscent of the ``mini-rebrightenings" recorded in the Kepler data of V585 Lyr \citep{Kato_2013_V585} and the TESS data of V844 Her \citep{Kato_2022_V844}.}
    \label{fig:Mini}
\end{figure}

\section{Analysis} \label{sec:analysis}

\subsection{Survey photometry} \label{subsec:survey}
    
    Figure~\ref{fig:longterm_light_curve} plots survey photometry of K2BS5 obtained by the Catalina Real-Time Transient Survey \citep[CRTS;][]{drake09}, the All-Sky Automated Survey for Supernovae \citep[ASAS-SN;][]{shappee, kochanek}, Asteroid Terrestrial-impact Last Alert System \citep[ATLAS; ][]{atlas}, and Gaia. Perhaps the most striking property of these data is the absence of outbursts; only a single undisputable outburst is present (in 2019), although there is a possible second outburst near the beginning of 2006. Owing to K2BS5's proximity to the ecliptic, there are significant seasonal gaps that could conceal additional superoutbursts. Indeed, the \ktwo\ light curve, the baseline of which is indicated in Fig.~\ref{fig:longterm_light_curve}, recorded a superoutburst in 2017 during one of those gaps.
    
    The 2019 outburst lasted for $\sim2$~weeks and has the shape of a superoutburst. The ATLAS data reveal that when K2BS5 emerged from solar conjunction in 2019, it was $\sim0.3$~mag fainter than usual and remained so for the next two months, after which it entered a superoutburst.

\begin{figure*}
    \centering
    \includegraphics[width=\textwidth]{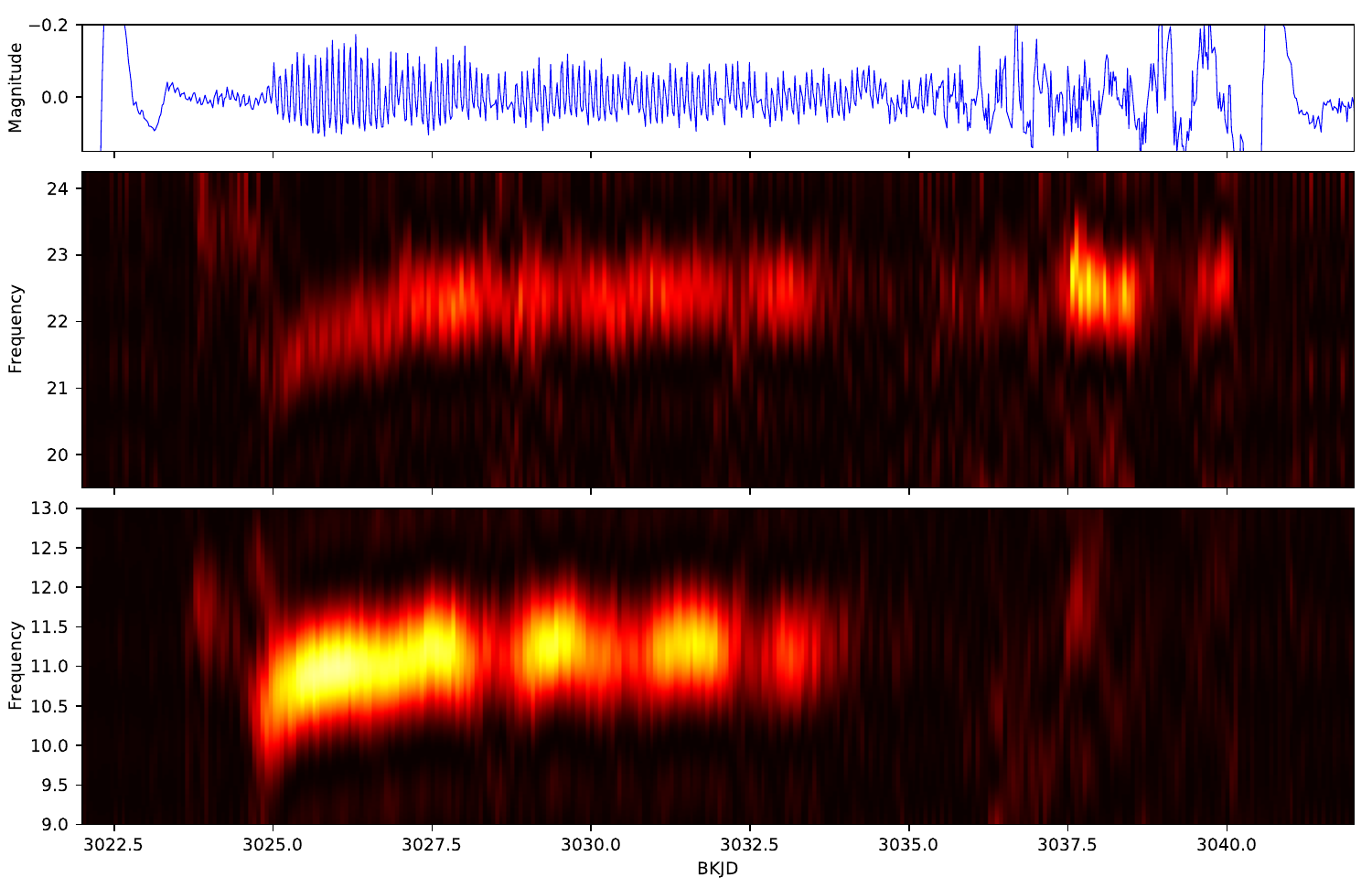}
    \caption{The shared-axis, detrended lightcurve (top) and time-resolved power spectrum (bottom) of K2BS5 starting just before the maximum of the superoutburst. Frequency units are cycles d$^{-1}$. Double-wave oscillations near the orbital period begin slightly before BKJD~3024 and transition rapidly into stage A superhumps, marked by the frequency drop on the power spectrum mid day 3024. Stage A is also relatively short lived, rising rapidly to higher frequency stage B. While the superhumps are not present during the \textbf{rebrightening event}, they do reappear just before it, between BKJD~3037 and BKJD~3039. Both the fundamental and second harmonic are visible on the power spectrum. Note: different intensity cuts were used in the second and third panels of the figure to improve signal visibility of the second harmonic.}
    \label{fig:Superoutburst 2}
\end{figure*}

\subsection{\ktwo\ Light Curve and Power Spectrum}

    In Figure \ref{fig:Superoutburst}, we show the \ktwo\ light curve of K2BS5, the most notable features of which are the superoutburst and a subsequent rebrightening event. The superoutburst occurs after at least three weeks of quiescence and begins with a steep rise of over 3~mag near BKJD = 3022, where BKJD is the Barycentric Kepler Julian Date, defined as BJD-2454833). After reaching its peak brightness near BKJD = 3024, the light curve begins to experience a slow fade, and large-amplitude superhumps appear. Near BKJD = 3033, the rate of fading increases dramatically. Notably, in the weeks following the superoutburst, K2BS5 never fully faded to its pre-superoutburst quiescent level, remaining $\sim0.3$~mag brighter after the superoutburst than it was before. As we noted in Sec.~\ref{subsec:survey}, the ATLAS observations of the suspected superoutburst in 2019 also show a $\sim0.3$~mag discontinuity in the quiescent brightness level before and after the superoutburst.

    The first rebrightening event was observed at BKJD=3040.3 for $\sim$ 2~d and showed a prominent superhump signal, while the second rebrightening occurred 20~d later. Unfortunately, the \ktwo\ campaign terminated while the second rebrightening was underway. Unlike the initial rebrightening, this second event is not accompanied by superhumps. Although the nature of the second event is somewhat ambiguous, both events appear to be causally related to the superoutburst. We base this inference on (1) the conspicuous absence of other outbursts of comparable amplitude in the long-term light curve (Fig.~\ref{fig:longterm_light_curve}) and (2) the fact that K2BS5 was still slightly brighter than its pre-quiescent brightness.
    % Additionally, low-amplitude oscillations with a period of $\sim$0.65~d appear during the post-superoutburst phase just before the first rebrightening. These $\sim$0.36~mag oscillations seen in Figure \ref{fig:Mini} are reminiscent of the ``mini-rebrightenings" recorded in the Kepler data of V585 Lyr \citep{Kato_2013_V585} and the TESS data of V844 Her \citep{Kato_2022_V844} and appear to continue between the first and second rebrightenings. 

    To better understand the changes in the light curve, we used {\tt astropy} \citep{Astropy} to create a two-dimensional Lomb-Scargle power spectrum \citep{LS_1976, LS_1982} with a sliding 0.5~d window. The lower panel of Fig.~\ref{fig:Superoutburst} presents the 2D power spectrum for the full dataset, while Fig.~\ref{fig:Superoutburst 2} shows an enlarged version of the 2D power spectrum during the superoutburst. The steep rise of the superoutburst at BKJD = 3024 coincides with the appearance of steady oscillations with a period of $\sim$ 2~h, nearly identical to the orbital period. This behavior is consistent with the expected behavior of ESH in WZ Sge systems---but, for reasons we set forth in Sec.~\ref{sec:discussion}, we do not classify them as such.

    After $\sim$1~day of these oscillations, the dominant frequency in the power spectrum quickly drops to a much lower frequency of 10.9~cycles/day, which we identify as the onset of Stage~A superhumps. This signal increases in frequency over the two days before leveling off around 11.2~cycles/day. The superhump power fades significantly near BKJD~3034 and reemerges 1.5~d before the rebrightening. The oscillations that redevelop just before the rebrightening are seen most strongly in the second harmonic of the superhump frequency. 

\begin{figure*}
    \centering
    \includegraphics[width=2.1\columnwidth]{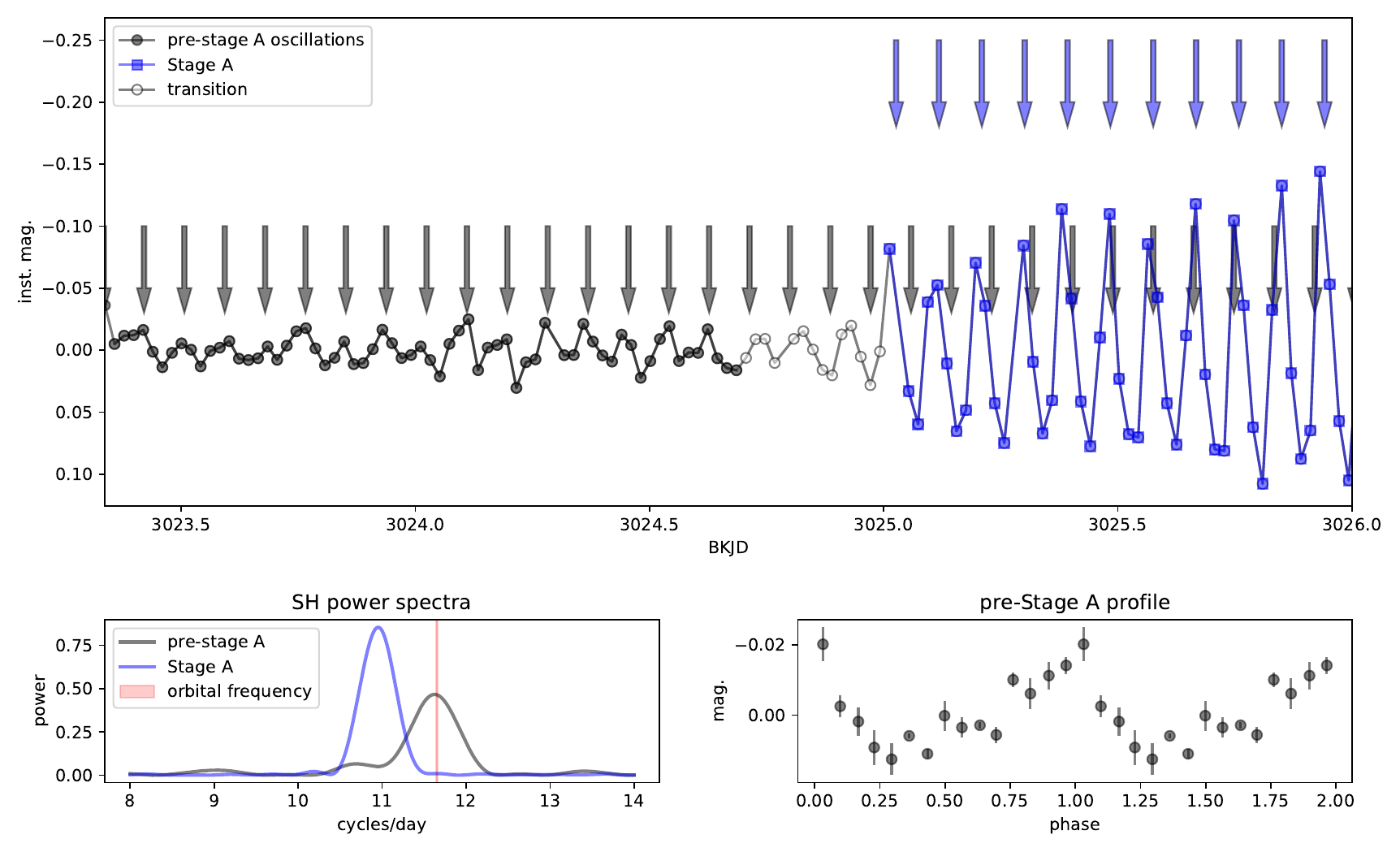}
    \caption{Top: Transition from double-wave oscillations to Stage A superhumps. The expected times of superhump maxima are indicated with black arrows (for the double-wave oscillations) and blue arrows (for Stage A superhumps). Both the double-wave oscillations and Stage A superhumps show stable, periodic maxima. Double-wave oscillations persisted for $\sim$15~cycles before transitioning into Stage~A superhumps in just several superhump cycles. Bottom left: Power spectra of the double-wave oscillations and Stage A superhumps. Bottom right: Phase-averaged profile of double-wave oscillations. }
    \label{fig:SH_transition}
\end{figure*}

    Fig.~\ref{fig:SH_transition} plots the detrended light curve near the peak of the superoutburst, and it reveals that there are $\sim$15~cycles of the double-wave oscillations before the appearance of Stage~A superhumps. This stage transition is rapid, occurring in just several superhump cycles.

%\newpage
\subsection{Extinction, Absolute Magnitude, and Superoutburst Amplitude}
\label{sec:extinction}
To estimate the absolute magnitude of K2BS5, we use DECam observations taken on January 30, 2020 \citep{DECam}. K2BS5 was imaged in 3 filters: g, r, and i. In IRAF, aperture photometry was performed on K2BS5 and multiple surrounding stars. We find the apparent magnitude of K2BS5 in quiescence to be $g = 18.01\pm 0.04$, $r = 17.54\pm 0.03$, and $i = 17.19\pm 0.03$.
Using the Gaia EDR3 parallax of K2BS5 in \citet{Bailer-Jones} and an reddening estimate of $E(g-r) = 0.29 \pm 0.02$ based on a 3D dust map modeled by \citet{green}, we find the absolute magnitude of K2BS5 to be $M_g = 9.36 \pm 0.15$, $M_r = 8.88 \pm 0.15$, and $M_i = 8.54 \pm 0.15$, calibrated in the Gaia-SDSS-PS1 Proper Motion Catalog \citep{Tian_2017}. After conversion to Gaia magnitudes \footnote{\url{https://gea.esac.esa.int} }, we find a luminosity of $G = 8.92 \pm 0.13$, slightly brighter than the average CV with an orbital period of 2 hours \citep{Abrahams}.

%https://gea.esac.esa.int/archive/documentation/GDR2/Data_processing/chap_cu5pho/sec_cu5pho_calibr/ssec_cu5pho_PhotTransf.html

As seen in Figure \ref{fig:Superoutburst}, the superoutburst begins with a steep rise in flux, corresponding to a magnitude increase of $\sim$3.3~mag as estimated directly from the \ktwo\ data. However, given the large aperture needed to mitigate the \ktwo\ drift and the blending of K2BS5 with nearby stars, it is likely that the \ktwo\ photometry is overestimating the quiescent brightness of K2BS5. To refine our estimate of the outburst amplitude, we first converted the flux measurement at the peak of the superoutburst to an r-band magnitude. Because the effects of contamination are minimal when K2BS5 is brightest, we can assume that this inferred maximum r-magnitude is accurate. However, during quiescence, the DECam images offer a more accurate measurement of K2BS5's brightness. Using these two measurements, we find the true amplitude of the outburst to be 3.8$\pm 0.05$~mag.

\subsection{LBT Spectrum \& Ground-Based Photometry}

\begin{figure*}
    \centering
    \includegraphics[width=\textwidth]{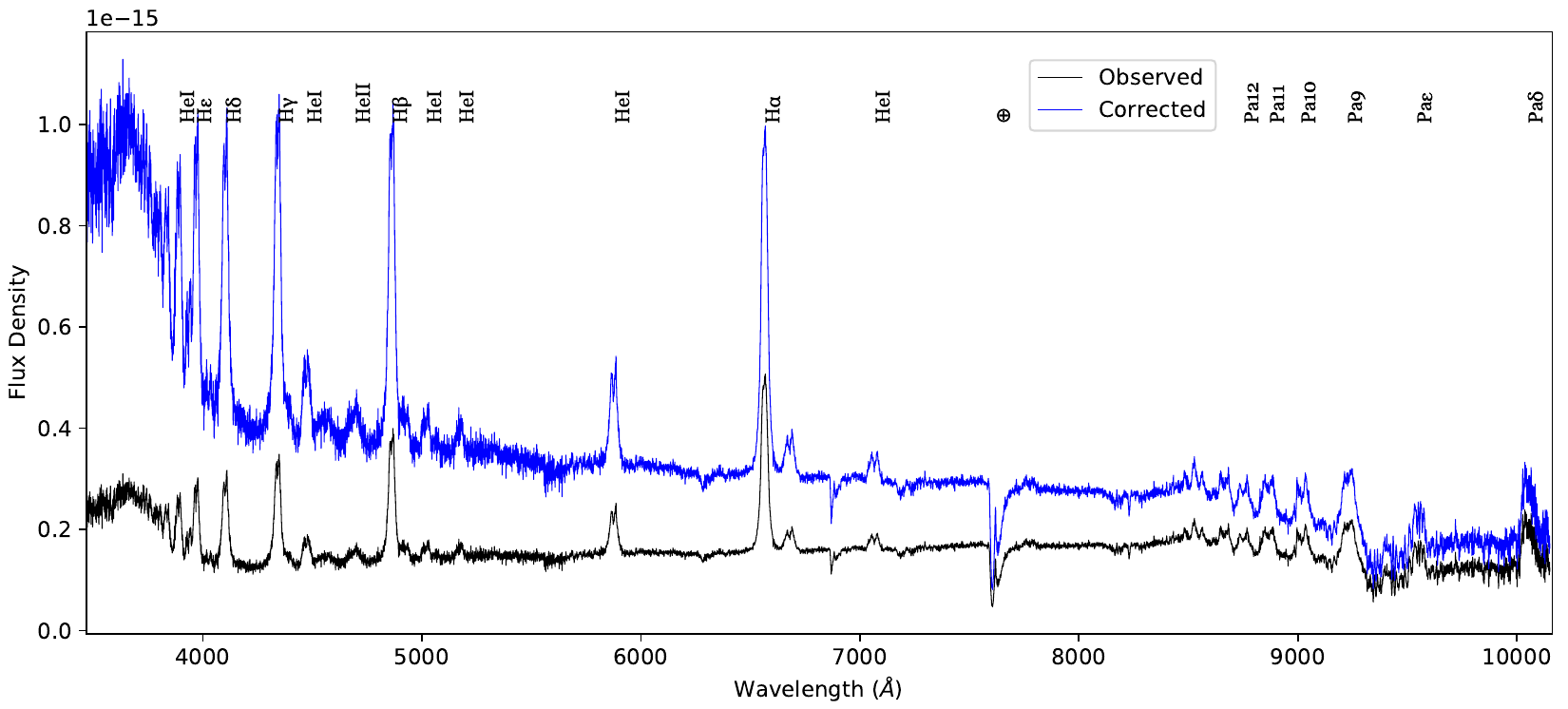}
    \caption{LBT quiescent spectrum of K2BS5 with no reddening correction (black line) and after correction for a reddening of $E(g-r) = 0.29$ mag (blue line). Wavelength units are given in angstroms and flux density units in erg~s$^{-1}$~cm$^{-2}$~\AA$^{-1}$. }
    \label{fig:Reduced Spectrum}
\end{figure*}

Obtained approximately 150 days after the peak of the 2019 superoutburst, the LBT spectrum of K2BS5 (Figure~\ref{fig:Reduced Spectrum}) shows broad, double-peaked hydrogen and helium emission features typical of a quiescent dwarf nova. The continuum is relatively flat except for a significant rise at the Balmer jump. After correcting for dust extinction as described in Sec.~\ref{sec:extinction}, the continuum rises slightly toward the blue, consistent with a disk dominated CV in quiescence. The full-width at half maximum (FWHM) of the H$_{\beta}$ emission line is 2020$\pm 20$~km$\;$s$^{-1}$. A weak He~II emission feature is seen that was not present in the \citet{thorstensen20} spectrum. 

The presence of He~II $\lambda4686$\AA\ in a CV spectrum can be attributed to high-temperature plasma or photoionization, conditions generally not present in quiescent DN systems. Strong He~II emission can be a sign of accretion onto a magnetic white dwarf, although here, the He~II line is not especially strong. The presence of a fast-spinning WD can be tested with high cadence optical photometry, which could detect the rotational period of the WD. Because \kepler's 30 min cadence is too slow to search for plausible spin periods, we analyzed the power spectrum of our comparatively fast-cadence SLKT observations for evidence of a short-period periodicity. We found no evidence of any such signal up to a frequency of 1300~cycles~d$^{-1}$. We conclude that there is no persuasive evidence that the WD is magnetized.

\section{Discussion} \label{sec:discussion}

\subsection{Mass Ratio}

Theory and observation show that the stellar mass ratio in SU~UMa and WZ~Sge binaries is related to the ratio between the binary orbital period and the superhump period. 

%{\bf The precession rate of the accretion disk over the can be expressed %in terms of the radius \textit{r} as follows \citep{Kato2013}:
%\begin{equation}
%    \frac{\omega_{pr}}{\omega_{orb}} = 
%    \frac{q}{\sqrt{1+q}} %\left(\frac{1}{4}\frac{1}{\sqrt{r}}b^{(1)}_{3/2}\right)
%\end{equation}
%Here, \textit{r} is expressed in units of the binary's separation, %$\omega_{orb}$ is the angular frequency of the binary, $q=M_2/M_1$, and %$\frac{1}{2}b^{(j)}_{s/2}$ is the Laplace coefficient \citep{Kato2013}, %defined as
%\begin{equation}
%    \frac{1}{2}b^{(j)}_{s/2}(r)=  \frac{1}{2\pi}\int_{0}^{2\pi} %\frac{\cos(j\phi)d\phi}{(1+r^2-2r\cos\phi)^{s/2}}
%\end{equation}}

%We derive the superhump period, and then infer $\epsilon$*, which is the %apsidal precession rate of the accretion disk over the orbital frequency %\citep{Kato2013}: 

%\begin{equation}
%\epsilon^* = \frac{\omega_{pr}}{\omega_{orb}} = 1 - %\frac{P_{orb}}{P_{SH}}
%\label{eqn:epsilon}
%\end{equation}

\begin{figure*}
    \centering
    \includegraphics[width =0.7\textwidth]{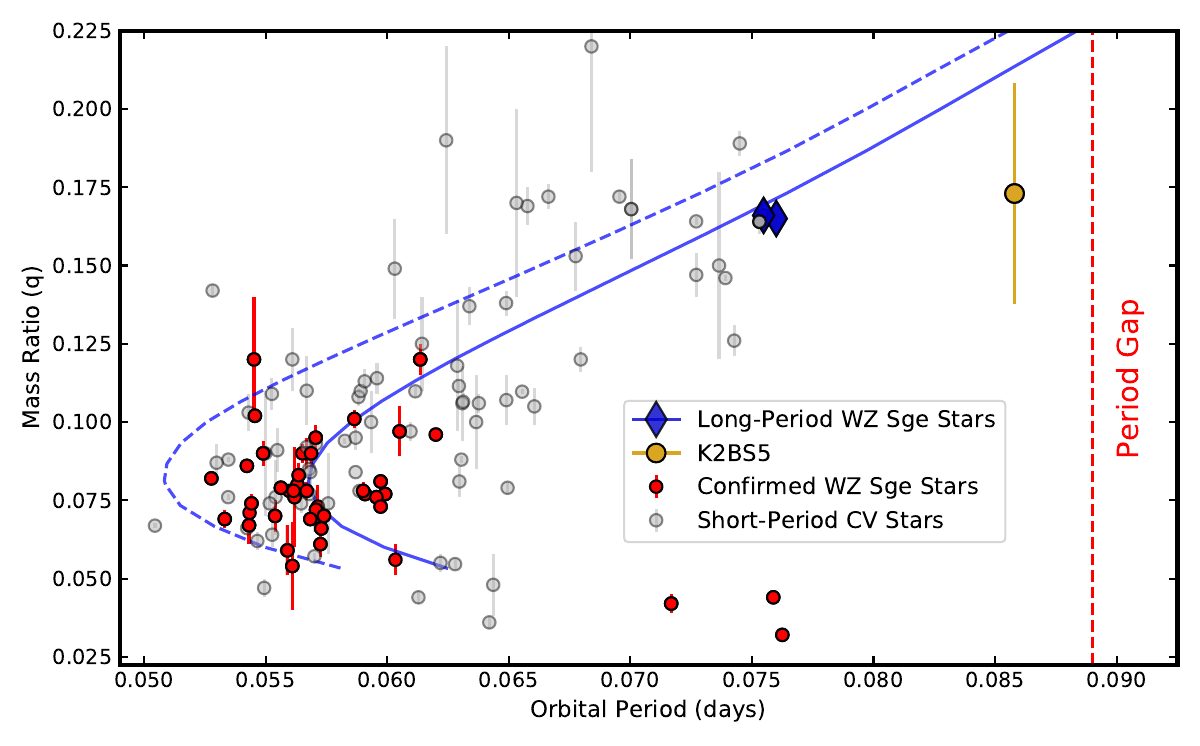}
    \caption{A distribution of the estimated mass ratio ($q$) versus binary orbital period of known WZ~Sge stars. The dashed blue line shows the standard CV evolutionary track from \citet{Knigge_2011}, while the solid blue line represents their optimal binary track. The dashed red line marks the short-period edge of the orbital period gap \citep{knigge06}. Confirmed WZ~Sge stars from \citet{Kato2015} and \citet{Kato_2022a} are labeled in red alongside short period CV stars from \citet{Kato_2022b} in grey. K2BS5 is shown in gold, while the long-period WZ~Sge systems RZ~Leo and ASASSN-16eg \citep{16eg} are plotted as blue diamonds.}
    \label{fig:WZ Sge Distribution}
\end{figure*}

Following the approach of \citet{Kato2013}, we derive the superhump period from Stage A, and then infer the value of $\epsilon$* from the following equation:
\begin{equation}
\epsilon^* = \frac{\omega_{pr}}{\omega_{orb}} = 1 - \frac{P_{orb}}{P_{SH}}
\label{eqn:epsilon}
\end{equation}
in which $\omega_{pr}$ represents the apsidal precession rate of the accretion disk, and $\omega_{orb}$ represents the orbital frequency \citep{Kato2013}. The orbital frequency is not discernible in the power spectrum, but it is already known from the spectroscopic study by \citet{thorstensen20}. Meanwhile, we measure a Stage~A frequency of 10.936$\pm 0.045$ cycles~d$^{-1}$ from the power spectrum.\footnote{We estimated the uncertainties by injecting into the light curve synthetic sinusoids with the same amplitudes as the superhumps; we then measured the resulting frequency in the power spectrum, calculated the error in frequency, and repeated the procedure using a different frequency. The standard deviations of the resulting distributions are the uncertainties for our superhump period measurements.} Combining the \citet{thorstensen20} orbital period with our measurement of the Stage~A period and following \citet{Kato2013},\footnote{Eq.~1 from \citet{Kato2013} contains a misprint 
\citep[described in][]{Kato_2022a},
and we used the corrected form of that equation in our calculation.} we determine the mass ratio of the system to be $q=0.173 \pm 0.035$. This ratio is typical of an SU UMa-type system and would be significantly higher than most known WZ Sge-type systems (Figure \ref{fig:WZ Sge Distribution}).  In WZ~Sge stars, \citet{Osaki_2002} estimate the upper limit to be $q$ $\leq$ 0.08, and according to \citet{Kato2015} most fall at or below $q = 0.06$. As the calculated mass ratio (q) falls well below the typical evolutionary track, we also calculated the value of q independently with the Stage~B frequency to verify our Stage~A measurements were not contaminated by Stage~B. From the Stage~B frequency of 11.160$\pm 0.089$ cycles~d$^{-1}$, and with the method provided by \citet{Kato_2022a}, we calculated the mass ratio of the system to be $q=0.175 \pm 0.025$. Thus, the independent calculations of $q$ are in excellent agreement. The system's modest divergence from the typical evolutionary track in Figure \ref{fig:WZ Sge Distribution} suggests the presence of a heavy white dwarf, which might also account for the presence of He II in Figure \ref{fig:Reduced Spectrum}.

\subsection{The First Rebrightening}

The power spectrum of leading up to the first rebrightening shows significant power at the superhump frequency and its second harmonic (Fig.~\ref{fig:Superoutburst}), which suggests that the disk remained tidally deformed even after the superoutburst faded. 
One possible explanation for the enhancement of the second superhump harmonic comes from simulations and observations by \citet{wood}. They showed that at the conclusion of a superoutburst, the interaction between the accretion stream and the outer disk can boost power at the second harmonic of the superhump frequency because the relative depth of the stream-disk hotspot in the WD's gravitational potential varies across the superhump cycle when the rim of the outer disk is eccentric \citep{wood}. 
% This provides support for the model of \citet{Hellier_Paper}, who proposed that the thermal and tidal instabilities that drive SU~UMa stars can become decoupled, in the sense that the thermal instability can subside even while the disk is still eccentric. 

We also see several ``mini-rebrightenings," each lasting for $\sim$0.3~d with an amplitude of $\sim$0.3-0.4~mag, in the trough immediately before the first rebrightening. These mini-rebrightenings seen in Fig.~\ref{fig:Mini} might be identical to the similarly named phenomenon observed in V585~Lyr by \citet{v585lyr}. The V585~Lyr mini-rebrightenings were also observed during a dip in the light curve immediately preceding a rebrightening, and their amplitudes, recurrence intervals, and durations were all comparable to what we see in K2BS5. The major difference is that the mini-rebrightenings in K2SB5 are comparatively ill-defined, with only two or three visible. This is far fewer than the nine very obvious mini-rebrightenings in Fig.~7 of \citet{v585lyr}. A similar phenomenon was also observed by \citet{Kato_2022_V844} in V844~Her.

\subsection{An SU~UMa system with some properties of WZ~Sge stars}

The pre-Stage-A oscillations near the superoutburst maximum are the most intriguing feature in the light curve, as they resemble ESH, the presence of which is often considered a defining quality of WZ~Sge systems. As summarized in \citet{Kato2015}, ESH are low-amplitude, double-peaked modulations that occur within $\sim0.1\%$ of the binary orbital period; they appear near the superoutburst maximum and always precede ordinary superhumps. On one hand, the pre-Stage-A oscillations in K2BS5 appear when ESH would be expected, are consistent with the known orbital period, and have a photometric profile compatible with the compilation in Figure~11 of \citet{Kato2015}; moreover, their peak-to-peak amplitude of $\sim0.04$~mag is in excellent agreement with the histogram of ESH amplitudes in Figure~15 of \citet{Kato2015}. However, the period of the oscillations is too uncertain to establish that it matches the \citet{thorstensen20} orbital period to within $\sim0.1\%$, as is required of ESH \citep{Kato2015}. As a result, we do not claim these oscillations to be ESH.\footnote{The large uncertainty of the photometric period is the result of two factors: the short duration of the pre-Stage-A oscillations ($\sim$15 cycles) and the low cadence of \kepler\ (which precludes us from using an O$-$C analysis of the maxima to more precisely measure their period). In WZ Sge systems, ESH often last for significantly longer, which facilitates the determination of a highly precise orbital period.}

Although it might be tempting to dismiss the pre-Stage-A oscillations as a transient peculiarity of just one dwarf nova, \citet{Kato_2022_V844} reported the presence of an apparently similar phenomenon in TESS observations of the SU~UMa-type dwarf nova V844~Her. \citet{Kato_2022_V844} noted that it is unclear as to whether double-waved oscillations are a general feature of SU~UMa systems and cautioned that this phenomenon should not be confused with ESH; however, that study did not explain how to distinguish the two using photometry alone. Considering the recent recognition of long-period WZ~Sge stars \citep[][and references therein]{16eg}, this point would benefit from elaboration, as the similarities between the two phenomena are sufficiently close that it can complicate classifications of WZ~Sge systems based solely on time-series photometry.

A full consideration of the criteria of WZ~Sge systems provides considerable evidence against K2BS5 being a WZ~Sge system, despite several similarities. In addition to ESH at the beginning of the superoutburst, WZ~Sge stars are generally characterized by several additional observational properties \citep{Kato2015}:
\begin{itemize}

\item one or more rebrightening events at the end of the superoutburst.

\item absence of a distinct precursor outburst \footnote{ The absence of a precursor outburst is sometimes observed in SU~UMa systems and is therefore not an exclusive property of WZ Sge stars, as seen in Case B of \citet{OM2003}}.

\item extremely long supercycles, defined as the average time between superoutbursts, with a minimum measured duration of 4 years \citep{Kato2015}.

\item large outburst amplitudes that typically exceed 7~mag.

\item orbital periods that are less than 0.065~d.

\item a mass ratio, $q$, generally less than 0.1.

\end{itemize}

With K2BS5, it is evident from Fig.~\ref{fig:Superoutburst} that there is no distinct precursor to the superoutburst, which is a property of WZ~Sge systems. The interval between supercycles, however, is a bit more ambiguous; there are large seasonal gaps in the ground-based survey photometry, and the K2 superoutburst occured during one of them. The interval between the only observed superoutbursts (in April 2017 and September 2019) suggests a supercycle of roughly 2.5 years (Figures \ref{fig:longterm_light_curve} and \ref{fig:Superoutburst}). This interval would be rather short for a typical WZ~Sge star, the supercycles of which typically range from 4 years to upwards of 30 years, with a majority of these systems having recurrence times shorter than $\sim$20 years. Nonetheless, Table~1 in \citet{16eg} lists two candidate long-period WZ~Sge systems, BC~UMa \footnote{As \citet{Maehara} and \citet{16eg} discuss, BC~UMa might be an intermediate object between typical WZ~Sge systems and SU~UMa systems, so it is not clear whether BC~UMa's properties can be generalized to long-period WZ~Sge objects. Indeed, when \citet{16eg} compared the supercycles of candidate long-period WZ~Sge stars, they excluded BC~UMa for this reason.} and V1251~Cyg, whose supercycles can be as short as 2 or 3 years, respectively. 
    
Another basic characteristic of WZ Sge-type systems is the rarity of normal outbursts during the periods between superoutbursts. The long-term light curve in Fig.~\ref{fig:longterm_light_curve} is compatible with this criterion, as K2BS5 appears to experience superoutbursts almost exclusively.

The argument for a long-period WZ~Sge interpretation of K2BS5 begins to fall apart on other grounds. In particular, long-period WZ~Sge systems are hypothesized to have unusually low mass-transfer rates at a given orbital period, which enables the outer disk to expand unusually far \citep{16eg}. Several different lines of evidence suggest that K2BS5's mass-transfer rate is too high to be in this regime. First, as we noted earlier, its absolute magnitude is slightly brighter than CVs of comparable orbital period \cite{Abrahams}. Furthermore, the presence of He~II $\lambda$4686\AA\ in the LBT spectrum underscores that the mass-transfer rate is not especially low. The \citet{16eg} mechanism for producing long-period WZ~Sge stars is therefore inapplicable to K2BS5.

Another argument against interpreting K2BS5 as a long-period WZ~Sge system is that ESH are detectable only above binary inclinations of $i \gtrsim40^{\circ}$ \citep{Kato2015, Kato_2022b}. Thus, if the pre-Stage-A oscillations were ESH, we would expect to detect the binary orbital period in the quiescent light curve. The absence of the orbital frequency in the quiescent power spectrum is consistent with a low orbital inclination. 

The superoutburst amplitude (3.8$\pm 0.05$~mag; see Sec.~\ref{sec:extinction}) is probably the most blatant observational dissimilarity with long-period WZ~Sge systems. In their Section 3.3, \citet{Kato2015} reported that 75\% of known WZ~Sge systems exhibit an outburst of at least 6.9~mag, with a median value of 7.7~mag. Moreover their Figure~3, which presents a histogram of the superoutburst amplitudes of WZ~Sge stars, only extends down to 5~magnitudes, which only underscores how extraordinarily low K2BS5's amplitude is when compared to typical WZ~Sge systems.

On balance, K2BS5 is best characterized as an SU~UMa system that shows some deceptive observational similarities with WZ~Sge systems. The nature of the pre-stage-A oscillations is unclear, but given the presence of a similar phenomenon of unknown origin in V844~Her \citep{Kato_2022_V844}, future \kepler- and TESS-based studies of SU~UMa systems should search for this phenomenon to ascertain both its prevalence and physical origin.

\section{Conclusion} \label{sec:conclusion}

K2BS5 is an SU~UMa-type dwarf nova with infrequent superoutbursts, no observed normal outbursts, and a mass ratio of $q = 0.173\pm 0.035$. Its most notable property is the short-lived appearance of double-peaked oscillations near the peak of the superoutburst, prior to the emergence of ordinary superhumps. The period of these oscillations agrees (within the errors) with the spectroscopic orbital period from \citet{thorstensen20} and is significantly shorter than the periods of the subsequent ordinary superhumps. Observationally, this phenomenon could easily mimic the early superhumps observed in WZ~Sge systems, but their period cannot be measured with sufficient precision to test whether they agree with the orbital period to within $0.1\%$ (a prerequisite for classifying them as early superhumps).

\begin{acknowledgments}

We thank the referee, Taichi Kato, for a swift and detailed report that reshaped our interpretation of the data and significantly improved this paper.

\end{acknowledgments}

    \software{ {\tt astropy} \citep{Astropy}
    {\tt lightkurve} \citep{Lightkurve}}
    
\bibliography{main.bbl}

\end{document}